\documentclass{PoS}

\title{Tracker and Calorimeter Performance for the Identification of Hadronic Tau Lepton Decays in ATLAS}

\ShortTitle{Performance of Tau Lepton Identification in ATLAS}

\author{\speaker{Stanley T. Lai}\thanks{on behalf of the ATLAS Collaboration.}\\
        Albert-Ludwigs-Universit\"{a}t Freiburg\\
        E-mail: \email{stan.lai@cern.ch}}

\abstract{

Tau leptons play an important role in the physics program in ATLAS. They can be used not only in 
searches for new phenomena like the Higgs boson or Supersymmetry, or for electroweak measurements but also 
in detector related studies like the determination of the missing transverse energy scale. Identifying 
hadronically decaying tau leptons requires good understanding of the detector performance, combining information from 
calorimeter and tracking detectors. The current status of the tau reconstruction and 
identification with the ATLAS detector is presented. The identification efficiencies are measured with 
$W\rightarrow\tau\nu$ events, and found to be consistent with the prediction from Monte Carlo simulations. 
}

\FullConference{The 2011 Europhysics Conference on High Energy Physics-HEP 2011,\\
		July 21-27, 2011\\
		Grenoble, Rh\^{o}ne-Alpes France}

\begin{document}

\section{Introduction}

The tau lepton, with a mass of $1776.82\pm0.16$~MeV~\cite{PDG}, is the only lepton heavy 
enough to decay both leptonically and hadronically.  It decays approximately 65\% of the time to 
one or more hadrons and 35\% of the time leptonically.  The reconstruction and identification of 
tau leptons are important in many searches for new phenomena~\cite{CSC}.
%, and they can appear in final states in the production of Higgs bosons, supersymmetric
% (SUSY) particles, and other particles not described by the Standard Model.  
Standard Model processes such as $W\rightarrow\tau\nu$, $Z\rightarrow\tau\tau$ boson production can 
also result in signatures with tau leptons, which can be used to 
measure key quantities such as the tau lepton identification efficiency.

A challenge in identifying hadronic tau decays ($\tau_\mathrm{had}$) 
is to distinguish them from hadronic jets.
%which are produced in processes with very large cross-sections.  
However, $\tau_\mathrm{had}$ leptons possess 
certain properties that can be used to differentiate them from jets.  
They usually decay into one (1-prong) or three (3-prong) charged particles 
and their decay products are well collimated.
%with an invariant mass less than $m_{\tau}$.  
The tau lepton proper lifetime is 87~$\mu$m, leading to decay 
vertices that can be resolved in the silicon tracker from the primary interaction vertex.

\section{Tau Reconstruction and Identification}

%The inner detector (tracker) and the calorimeter are crucial for the performance of the tau reconstruction
%algorithm~\cite{DetPap}.  The tracker consists of the Pixel Detector, the Semi-Conductor Tracker (SCT), and the
%Transition Radiation Tracker (TRT).  The Pixel Detector and SCT cover a region $|\eta| < 2.5$ in pseudorapidity,
%while the TRT covers a region out to $|\eta| < 2.0$.  The electromagnetic (EM) calorimeter alternates layers of
%liquid argon (LAr) as the active material, and lead as the absorber, covering out to $|\eta| < 3.2$.  The hadronic 
%calorimeter uses steel and scintillating tiles in the barrel region ($|\eta| < 1.7$), while the end-caps use LAr 
%as the active material and copper as the absorber ($1.5 < |\eta| < 3.2$).

In ATLAS~\cite{DetPap}, hadronically decaying tau candidates~\cite{taureco} are seeded by anti-$k_\mathrm{t}$ 
jets~\cite{antikt} with distance parameter $R=0.4$ satisfying $p_\mathrm{T}>10$ GeV and $|\eta|<2.5$ built 
from three dimensional clusters of calorimeter cells~\cite{topocluster}.
Tracks reconstructed with $p_T > 1$ GeV and within $\Delta R < 0.2$ of the jet seed are associated to the tau 
candidate if they satisfy cuts on the impact parameter and minimum silicon hit criteria.
%satisfy the following criteria:
%\begin{itemize}\addtolength{\itemsep}{-0.5\baselineskip}
%\item Number of pixel hits $\geq$ 2,
%\item Number of pixel hits + number of SCT hits $\geq$ 7,
%\item $|d_0| < 1.0~\mathrm{mm}$ and $|z_0 \sin{\theta}| < 1.5~\mathrm{mm}$.
%\end{itemize}
%where $d_0$ ($z_0$) is the transverse (longitudinal) impact parameter of the track.

The energy scale of tau candidates is determined using a two-step process.  In the first step,
local hadron calibration~\cite{LCcalib} is applied to clusters
within a radius of $\Delta R < 0.2$ of the barycentre.
%This weights the cell energies of the clusters, based on local properties of the clusters, in order to correct for 
%the non-compensation of the calorimeters, and for energy deposits outside reconstructed clusters and in 
%uninstrumented regions.  
The resultant energy from the sum of cluster four-vectors is used for the second step of the calibration.  
In this step, an additional correction on this energy is applied, based on 
Monte Carlo (MC) studies of processes involving hadronic tau decays, to obtain the fully calibrated 
tau candidate energy.  Uncertainties on the energy scale are determined by comparing 
the calibrated energy in different MC simulation samples, with realistic variations
of conditions such as the hadronic shower model and dead material modelling~\cite{taureco}.

\begin{figure}
    \centering
    \includegraphics[width=0.45\textwidth]{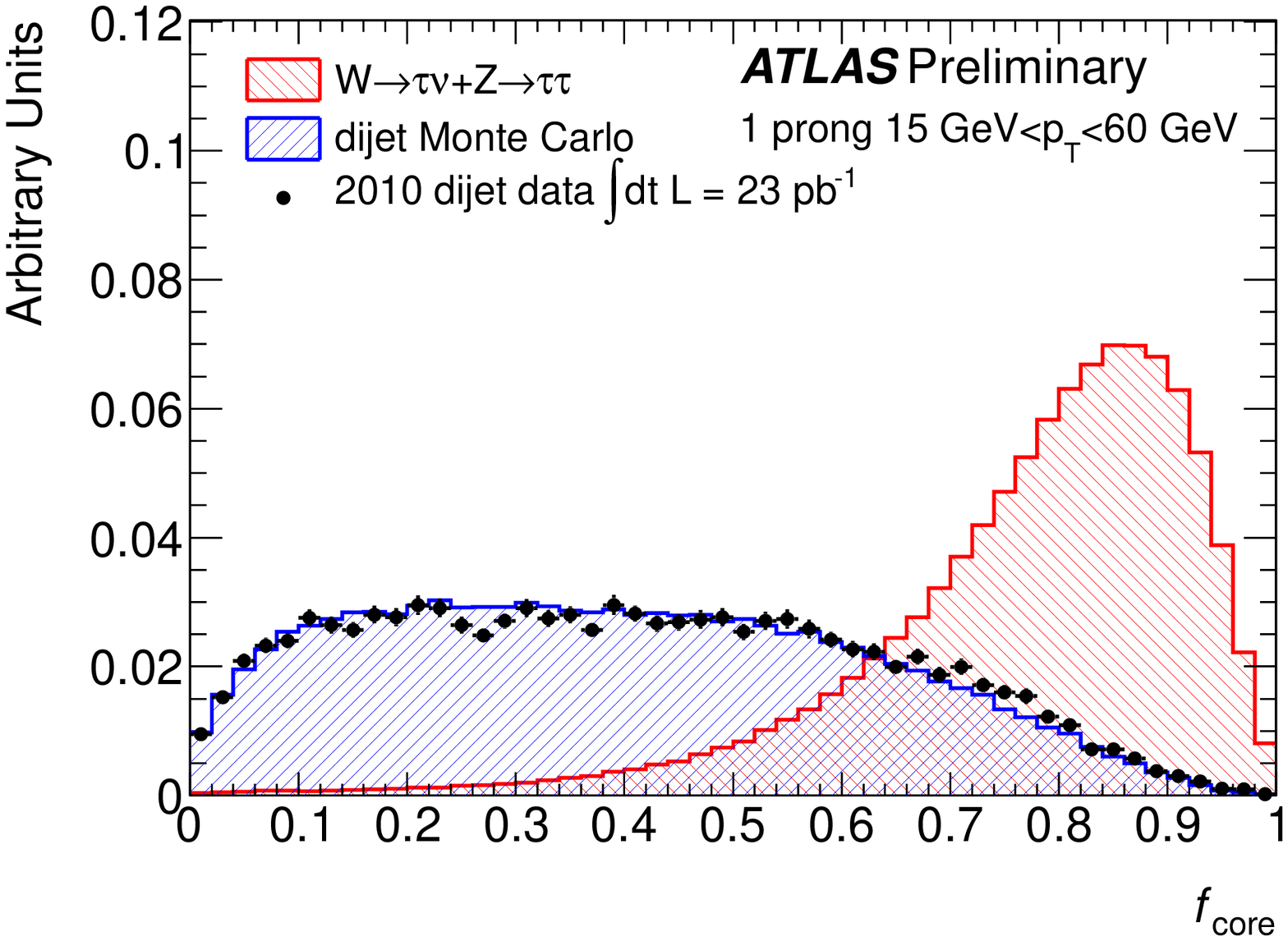}
    \includegraphics[width=0.45\textwidth]{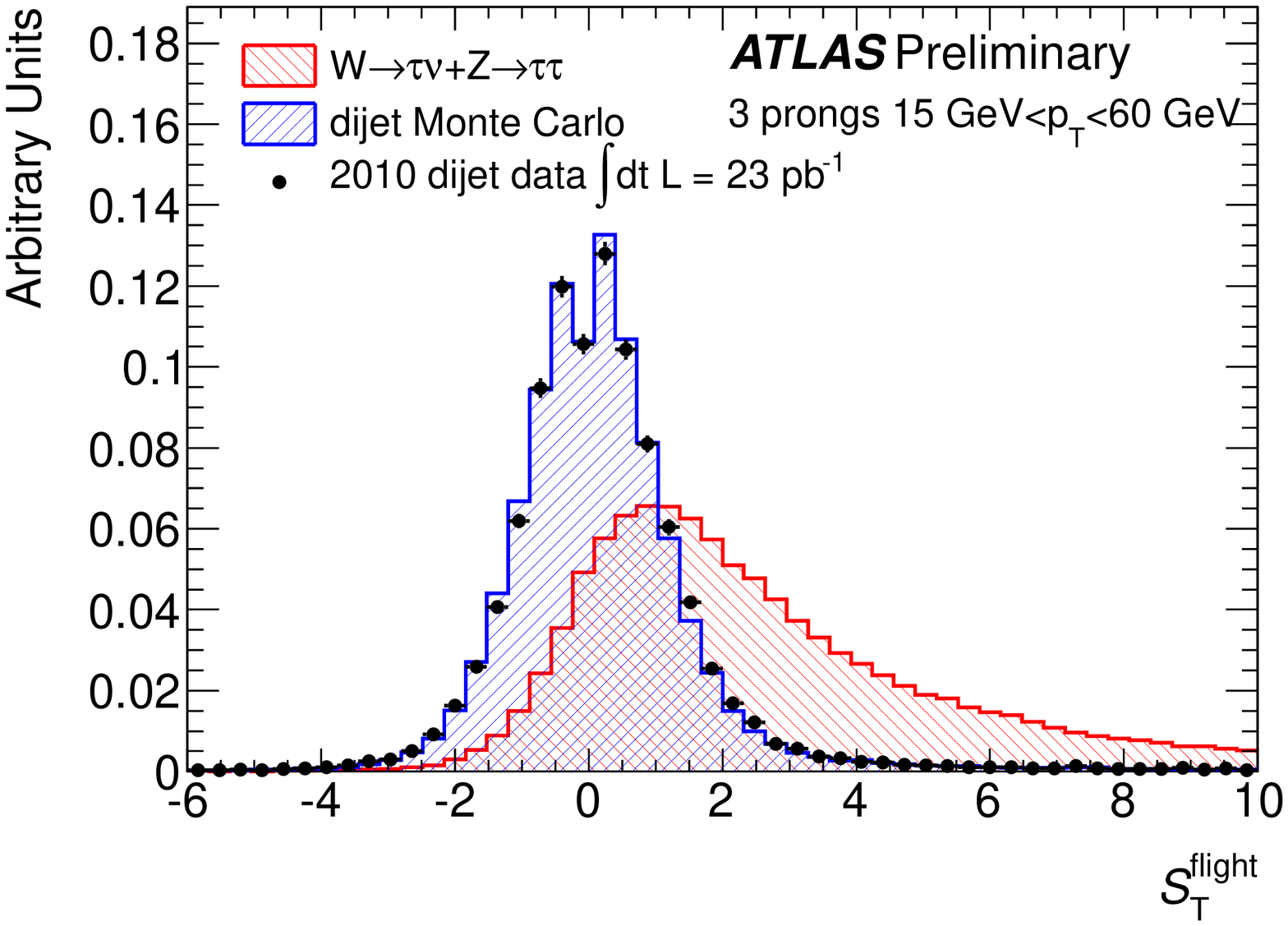}
        
    \caption{
    Two discriminating variables used for the tau identification: $f_\mathrm{core}$ and $S_\mathrm{T}^\mathrm{flight}$~\cite{taureco}. 
    The filled histograms are from MC simulation, the points are data with a dijet selection.
     \label{fig:IDvars}}
\end{figure}

Discrimination against background candidates from jets and electrons is provided in a separate identification step.
Identification variables are reconstructed from tau candidates, based on tracker and calorimeter 
information~\cite{taureco}.  Examples of such identification variables include: the core energy fraction
($f_\mathrm{core}$), the ratio of energies at the electromagnetic scale deposited within 
$\Delta R < 0.1$ and $\Delta R < 0.4$ of the tau candidate; and the transverse flight path 
significance ($S_\mathrm{T}^\mathrm{flight}$), 
the transverse decay length significance of the reconstructed vertex of multi-prong candidates.  
Distributions of $f_\mathrm{core}$ for 1-track candidates and 
$S_\mathrm{T}^\mathrm{flight}$ for 3-track candidates are shown in Figure~\ref{fig:IDvars}, for 
both signal (in simulated $W\rightarrow\tau\nu$ and $Z\rightarrow\tau\tau$ events) and 
jet background (from a dijet selection) candidates.

\begin{figure}
    \centering
    \includegraphics[width=0.45\textwidth]{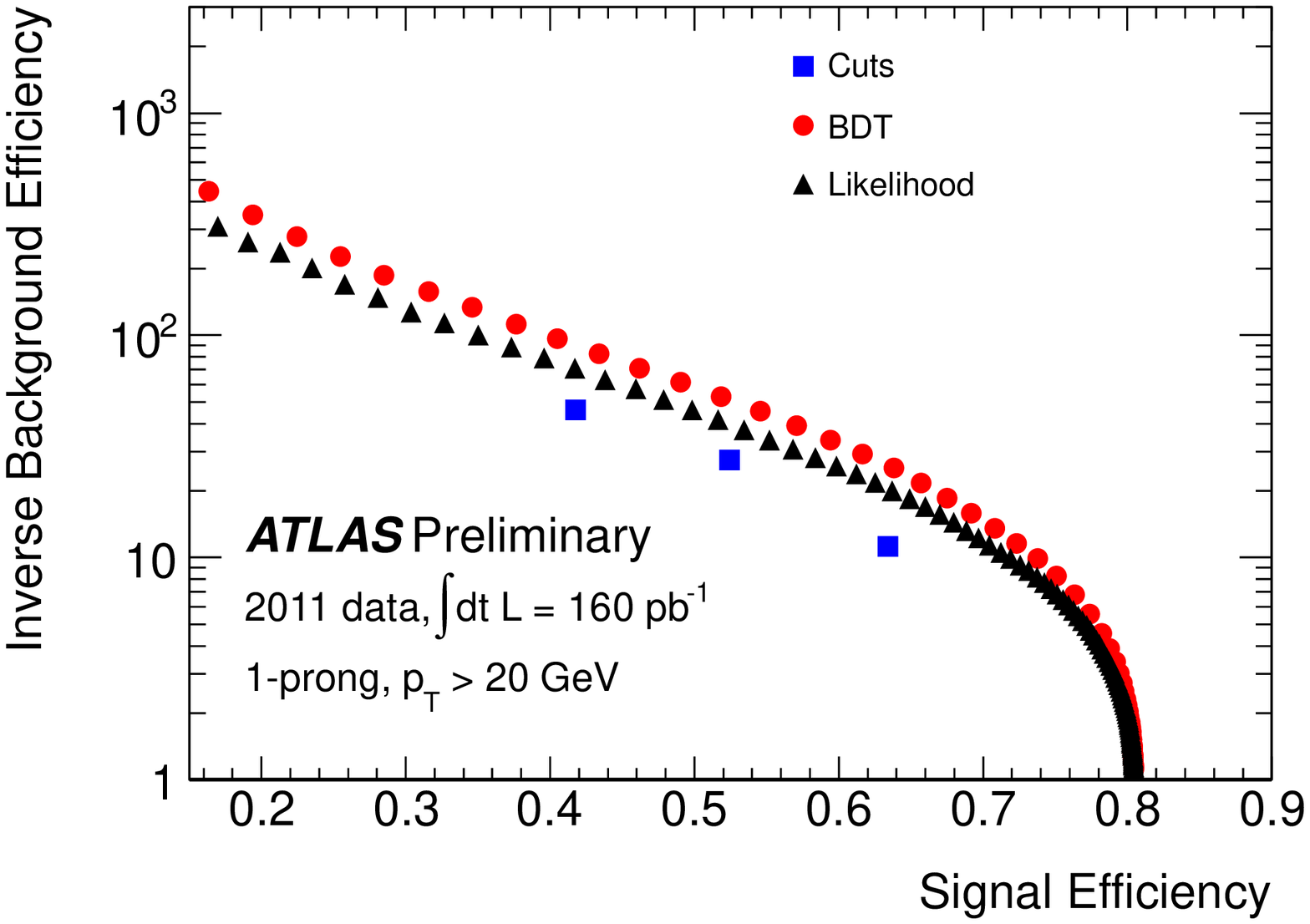}
    \includegraphics[width=0.45\textwidth]{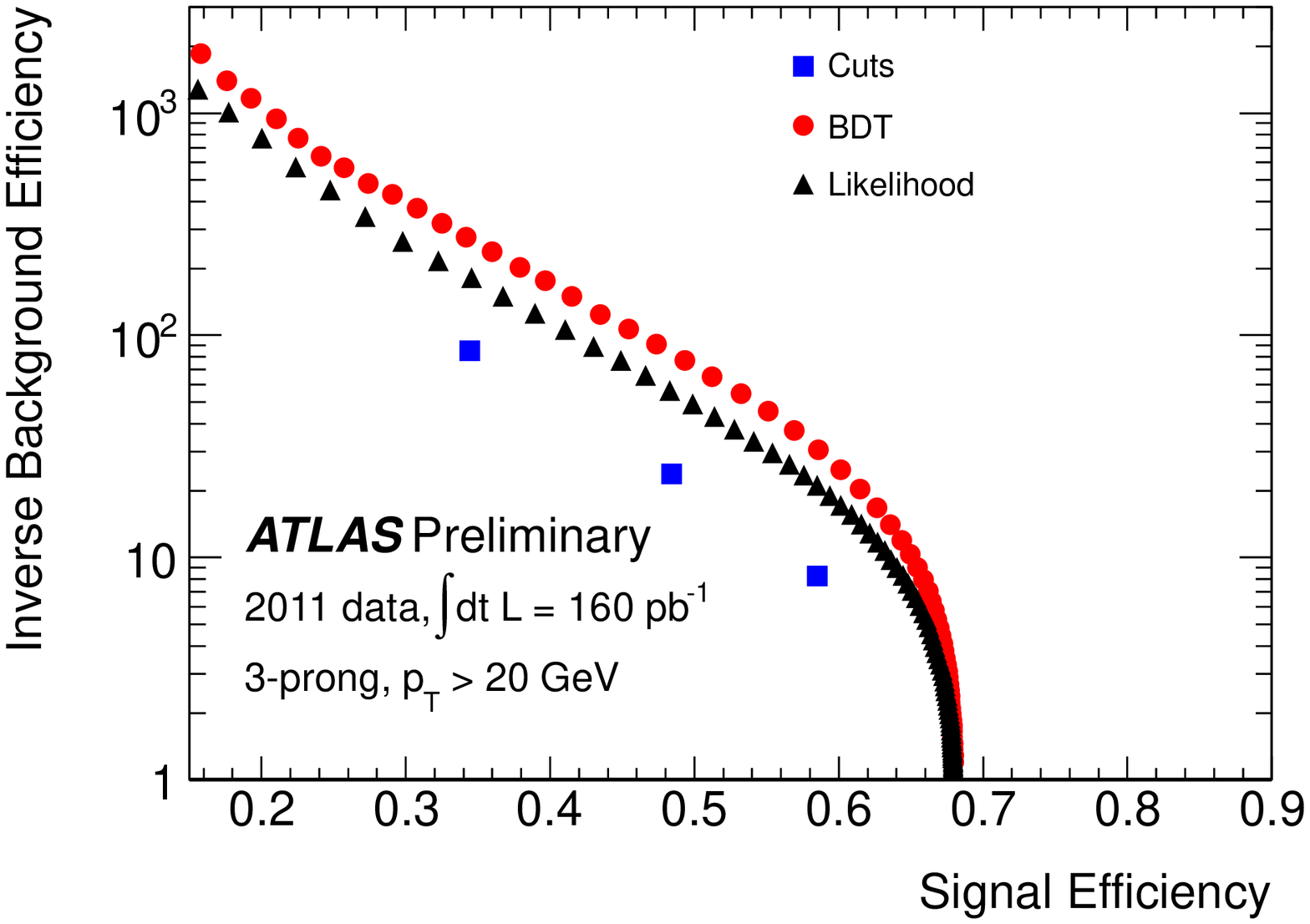}
        
    \caption{
    Inverse background efficiency in dijet data as a function of signal efficiency 
    in $W\rightarrow\tau\nu$ and $Z\rightarrow\tau\tau$ MC events for all 
    discriminants on 1-track and 3-track candidates~\cite{TauWiki}.
     \label{fig:Performance}}
\end{figure}

The identification variables are combined into discriminants to suppress background candidates from 
jets and electrons.  ATLAS has developed three such discriminants: a cut-based discriminant, 
a projective likelihood, and a boosted decision tree~\cite{taureco}.  The performance of these 
discriminants can be evaluated by plotting the rejection (inverse efficiency) for jets in a dijet 
selection against the efficiency for signal tau candidates, for both 1-track and 3-track candidates.  
This is shown in Figure~\ref{fig:Performance}.

\section{Tau Identification Efficiency Measurements}

\begin{figure}
    \centering
    \includegraphics[width=0.40\textwidth]{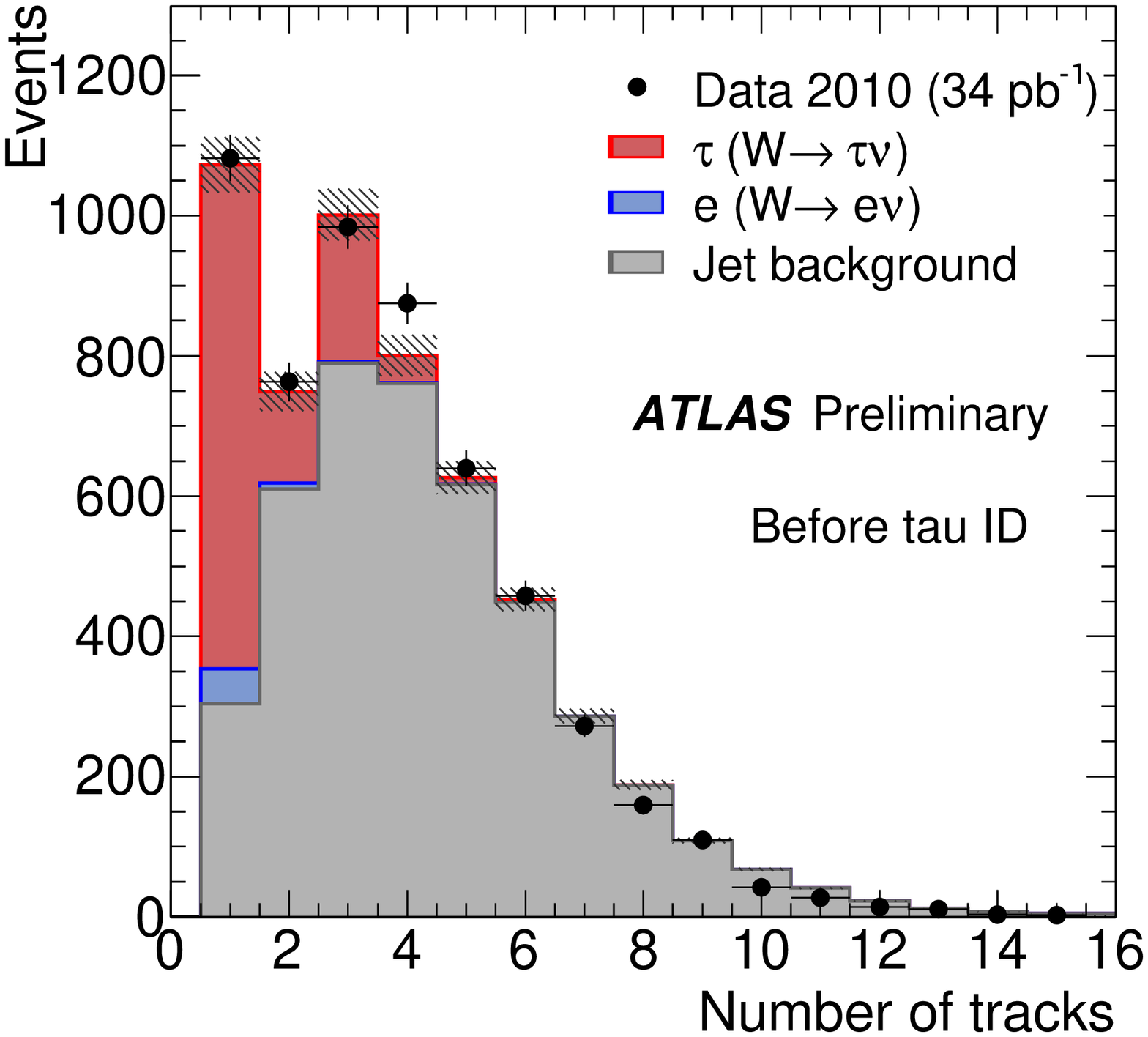}
    \includegraphics[width=0.40\textwidth]{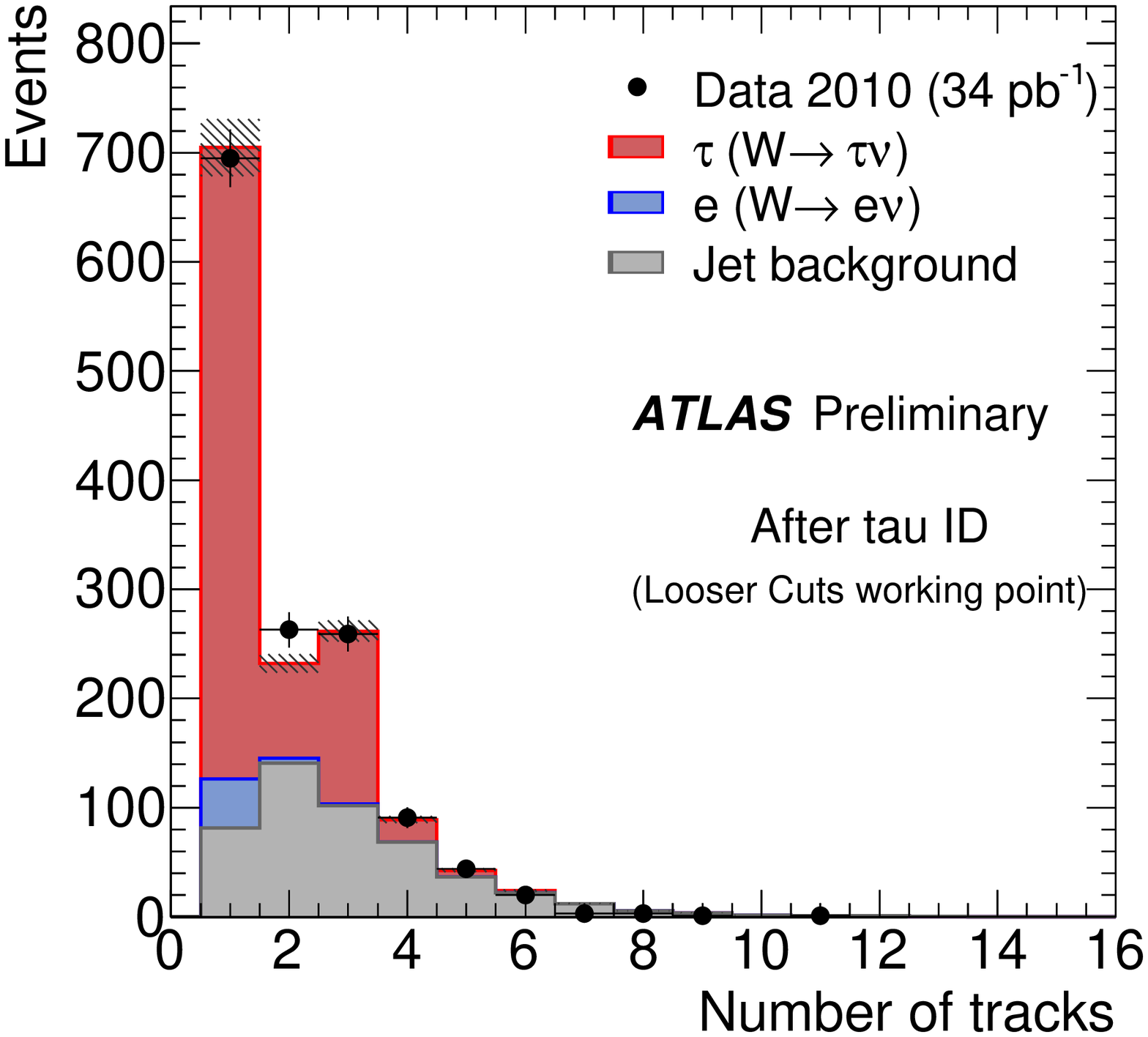}
        
    \caption{
    Track multiplicity before and after tau identification~\cite{WtaunuEff}. The hatching represents the systematic uncertainty. 
    The normalisation of the different processes is determined through a fit to the track multiplicity spectrum. 
     \label{fig:WtaunuTrack}}
\end{figure}

The identification efficiency for hadronic tau decays is measured with $W\rightarrow\tau\nu$ events~\cite{WtaunuEff}.  
In the tag \& probe method, events are selected with
$S_{E_\mathrm{T}^\mathrm{miss}} = E_\mathrm{T}^\mathrm{miss} / (0.5 \mbox{GeV}^{1/2}\sqrt{\Sigma E_\mathrm{T}}) > 6$,
where $E_\mathrm{T}^\mathrm{miss}$ is the missing transverse energy and $\Sigma E_\mathrm{T}$ is the scalar sum of 
cluster transverse energy.
A tau candidate that is well separated from the direction of the $E_\mathrm{T}^\mathrm{miss}$ is required,
and the tau track multiplicity is fitted before and after applying tau identification to extract the fraction of
$W\rightarrow\tau\nu$ events in the sample.  The track multiplicity templates for true $\tau_\mathrm{had}$ leptons 
are taken from $W\rightarrow\tau\nu$ MC events, while for jet candidates, a template from a jet-enriched 
control region $2 < S_{E_\mathrm{T}^\mathrm{miss}} < 4.5$ is used.  
The tau track multiplicity before and after tau identification is shown in Figure~\ref{fig:WtaunuTrack}.

This measurement is cross-checked with a second method that compares the observed $W\rightarrow\tau\nu$ 
event yield with the predicted $W\rightarrow\tau\nu$ event yield based on the measured
$W\rightarrow e\nu$ and $W\rightarrow \mu\nu$ cross-sections.
Both methods measure a tau identification efficiency in $W\rightarrow\tau\nu$ events that are 
consistent with the predicted efficiency from MC simulation~\cite{WtaunuEff}.

\newpage

\section{Summary}

ATLAS has developed a well-performing reconstruction algorithm to identify hadronic tau decays, enabling
various measurements and searches of physics processes with tau leptons in the final state.  MC predictions of the 
identification efficiency are shown to be consistent with measurements of the efficiency in 
$W\rightarrow\tau\nu$ data events.


\begin{thebibliography}{99}

\bibitem{PDG} 
K. Nakamura et al. (Particle Data Group), J. Phys. G 37, 075021 (2010). 

\bibitem{CSC}
The ATLAS Collaboration,  ``Expected Performance of the ATLAS Experiment, Detector, Trigger,
and Physics'', CERN-OPEN-2008-020, Geneva, (2008).

\bibitem{DetPap}
The ATLAS Collaboration, G. Aad {\it et al.}, ``The ATLAS Experiment at the CERN Large Hadron
Collider'', JINST 3 (2008) S08003.

\bibitem{taureco}
The ATLAS Collaboration, ``Reconstruction, Energy Calibration, and Identification of Hadronically Decaying Tau Leptons in the ATLAS Experiment", 
ATLAS-CONF-2011-077 (2011).
http://cdsweb.cern.ch/record/1353226.

\bibitem{antikt}
M. Cacciari {\it et al.}, ``The anti-$k_\mathrm{t}$ jet clustering algorithm", JHEP04 (2008) 063.

\bibitem{topocluster}
W. Lampl {\it et al.}, ``Calorimeter Clustering Algorithms: Description and Performance", ATL-LARG-PUB-2008-002 (2008).

\bibitem{LCcalib}
T. Barillari {\it et al.}, ``Local Hadronic Calibration", ATL-LARG-PUB-2009-001-2 (2009).
http://cdsweb.cern.ch/record/1112035.

\bibitem{TauWiki}
https://twiki.cern.ch/twiki/bin/view/AtlasPublic/TauPublicCollisionResults

\bibitem{WtaunuEff}
The ATLAS Collaboration, ``Measurement of Hadronic Tau Decay Identification Efficiency using $W\rightarrow\tau\nu$ events", ATLAS-CONF-2011-093 (2011).
\newline
http://cdsweb.cern.ch/record/1365728.


\end{thebibliography}
\end{document}